\documentclass[nohyper,notoc]{JHEP3}

\usepackage{amsmath,euscript,array,amssymb,cite,bm} 
\def\Tr{{\rm Tr}}

\def\det{{\rm det}}

\def\ui{{\underbar{i}}}
\def\uj{{\underbar{j}}}

\def\Dbarslash{\,\,{\raise.15ex\hbox{/}\mkern-12mu {\bar\D}}}
\def\Dslash{\,\,{\raise.15ex\hbox{/}\mkern-12mu \D}}
\def\delslash{\,\,{\raise.15ex\hbox{/}\mkern-9mu \partial}}
\def\delbarslash{\,\,{\raise.15ex\hbox{/}\mkern-9mu {\bar\partial}}}

\def\hf{{\textstyle{1\over2}}}

\def\D{{\cal D}}
\def\Dbarslash{\,\,{\raise.15ex\hbox{/}\mkern-12mu {\bar\D}}}
\def\delslash{\,\,{\raise.15ex\hbox{/}\mkern-9mu \partial}}
\def\Dslash{\,\,{\raise.15ex\hbox{/}\mkern-12mu \D}}

\def\={\, =\, }
\def\+{\, +\, }
\def\-{\, -\, }

\newcommand{\be}{\begin{equation}}
\newcommand{\ee}{\end{equation}}
\def\bea{\begin{eqnarray}}
\def\eea{\end{eqnarray}}

\def\p{\varphi}
\def\pt{\tilde\varphi}
\def\uno{\mbox{1 \kern-.59em {\rm l}}}

\title{Metastable SUSY breaking within the Standard Model}

\author{Steven A.~Abel and Valentin V.~Khoze\\
Institute for Particle Physics Phenomenology and Centre for Particle Theory, \\
University of Durham,
Durham, DH1 3LE, UK\\
{\tt s.a.abel@durham.ac.uk,}\,  
{\tt valya.khoze@durham.ac.uk}}

\abstract{We construct a supersymmetric version of the Standard Model which contains a long-lived
metastable vacuum. In this vacuum supersymmetry is broken and the electroweak symmetry is Higgsed, and 
we identify it with the physical ground state of the Standard Model. In our approach the 
metastable supersymmetry breaking (MSB) occurs directly in the $SU(2)_L \times U(1)_Y$ sector of the Standard Model;
it does not require a separate MSB sector and in this way it departs from the usual lore.
There is a direct link between the electroweak symmetry breaking
and the supersymmetry breaking in our model, both effects are induced by the same Higgs fields 
$\p_i, \pt_i.$ In order to generate sufficiently large
gluino masses we have to have strong coupling in the Higgs sector, $h \gg 1.$ 
Our model results in an extremely compact low-energy effective theory at the electroweak scale
with Higgs fields being very heavy, $M_{Higgs} \gg M_W$ and frozen at their vacuum expectation values.
}
 
\preprint{{\tt hep-ph/0701069}
\\  IPPP/07/01\\
}

\begin{document}

\section{Introduction}

Supersymmetry breaking in a long-lived metastable vacuum (MSB) 
discovered by Intriligator, Seiberg and Shih \cite{ISS}
is an exciting possibility for model building. MSB scenarios are based 
on models which contain supersymmetry-breaking metastable vacua in addition to the
supersymmetry-preserving (stable) ground states. The existence of supersymmetric vacua
in MSB models makes them much less constrained \cite{ISS} than the more traditionally
considered scenarios of dynamical supersymmetry breaking (DSB)  \cite{rev-DSB})
which contain only non-supersymmetric ground states.

Furthermore, MSB models are
at least as natural as the DSB framework. In particular
it has recently been shown that, in Intriligator-Seiberg-Shih (ISS) MSB
models, the early Universe would generically have been driven to the
metastable vacua by thermal effects \cite{ACJK,AJK}, and also that once
trapped there the lifetime for decay to the true (supersymmetric)
vacua is much longer than the age of the Universe for any reasonable
choice of parameters \cite{ISS,CFW,FKKMT,ACJK,AJK}, thus realising in 
an elegant way an early idea of Ellis {\em et al} \cite{ELR}. 
MSB therefore completes
the canon of supersymmetry breaking available to supersymmetric field
theory, so from a fundamental viewpoint it is certainly an interesting
development. On the other hand, the benefits of the MSB over DSB scenarios 
for phenomenological applications are less immediately obvious. 
If the MSB sector forms a hidden sector
then its phenomenological consequences are largely determined by the
method of mediation to the visible sector. Could one tell in practice
if a hidden sector were MSB or DSB?  

Intuitively it is clear that the more direct the mediation 
of SUSY breaking to the Standard Model is, the easier this
would be. In a series of recent papers \cite{DM,KOO,MN,CST,AS} the 
ISS-type models with metastable SUSY breaking were used to construct 
new, simple and calculable models of direct gauge mediation. In this set-up the ISS model forms a hidden sector
which breaks supersymmetry; this effect is then communicated to the Standard Model via gauge interactions. 
 
In this paper, we will pursue a different and more extreme approach which 
is to eliminate the need for a hidden sector at all: we will use the ISS
model as a basis for \emph{visible}
sector SUSY breaking. There is a more specific reason why ISS models
warrant a return to visible sector SUSY breaking, namely the fact
that their gauge groups are automatically Higgsed in the metastable minimum and
moreover the scale of Higgsing is of the same order as the scale of
SUSY breaking. Like O'Raifeartaigh models of old, a natural link between
gauge symmetry breaking and SUSY breaking comes for free in
MSB. 

We propose incorporating the ISS-type metastability directly into the
electroweak $SU(2)_L \times U(1)_Y$ sector of the MSSM in order to give a visible sector SUSY
breaking that is naturally linked to, and driven by, electroweak symmetry
breaking. In this proposal we are thinking of the MSSM-like theory
as a magnetic dual theory which is valid in the IR. Indeed just like
the ISS models, the model we put forward has a Landau pole in the
$SU(2)_{L}$ gauge coupling at some scale $\Lambda_{L}$. Above this
scale we will assume that there is an (unknown) electric theory which
takes over\footnote{We will see that the
scale where our theory will require a UV completion is comfortably very high, 
$\Lambda_{L}> M_{Pl}.$ The electric theory above $\Lambda_L$ can be a string or a field theory which 
is related to our magnetic theory below $\Lambda_L$ in some way, possibly involving
a generalisation of Seiberg duality \cite{Seiberg2}.}; however 
the form of the microscopic electric theory has no bearing
on phenomenology.
The model we propose needs very little extension beyond the conventional
MSSM, merely an ISS-like O'Raifeartaigh potential for the Higgs sector
and some extra generations of Higgs fields to cause $SU(2)_{L}$ to
become strongly coupled at $\Lambda_{L}$.

At first sight our proposal seems bound to fail because of two familiar
{}``no-go'' theorems. The first is the theorem by Nelson and Seiberg \cite{NS}
that SUSY breaking in a \emph{generic} theory requires an R-symmetry
(where \emph{generic} means that all operators that are allowed by
symmetries appear in the superpotential). This appears to exclude
the possibility of non-zero gaugino Majorana masses since they are inconsistent
with an unbroken R-symmetry.
A spontaneously broken R-symmetry on the other hand implies
a massless $R-$axion which is disallowed on cosmological grounds
\cite{BRP}. The second is a no-go theorem \cite{5917} coming from 
the well-known sum-rule $STr(M^{2})=0.$
This relation holds at tree-level even when SUSY is spontaneously
broken, and can be applied to differently charged fields independently,
so that for example it predicts $m_{\tilde{d}}^{2}+m_{\tilde{s}}^{2}+m_{\tilde{b}}^{2}\sim(5\mbox{GeV})^{2}$,
obviously completely at odds with experiment \cite{Kane}. 
To avoid this tree-level mass relation one has to generate SUSY
breaking terms of order $\sim$TeV at one-loop or higher. This implies
that the $F$-term vev responsible for SUSY breaking must be at least
$100$ TeV$^{2}$ as is the case in gauge mediated SUSY breaking
for example. Say the vev of the Higgs fields breaking electroweak
symmetry is $\mu\sim g_{2}^{-1}M_{W}$. Then since we want to induce
SUSY breaking and electroweak symmetry breaking with the same field
this implies $F=h\mu^{2}$ where $h$ is some coupling constant which
clearly has to be much greater than one. How can such large couplings
- and this is the essence of the problem - be consistent in a calculable
theory? We will show that both of these theorems are evaded by the
special properties of MSB models. 

The first crucial point is that, as pointed out by ISS, metastable
models do not \emph{have} to adhere to the Nelson-Seiberg theorem
because they have supersymmetric vacua, and indeed in ISS-type models
they violate it in an interesting way; the theory at the metastable
minimum resembles a standard O'Raifeartaigh model, SUSY is broken
and there is a global R-symmetry. However the global supersymmetric
minima are recovered by a nonperturbative dynamical term that is generated
by the $SU(2)_{L}$ gauge symmetry. The R-symmetry is anomalous
under this $SU(2)_{L}$ group and therefore the dynamical term does
not respect it. This strongly suggests that other sectors of the 
theory may dynamically produce R-symmetry violating operators as
well whilst leaving supersymmetry intact (as for example the magnetic
theory does in the supersymmetric minima). Depending on how the breaking
is mediated to the magnetic theory, one does not expect \emph{all}
possible operators to be generated \emph{at leading order}. The resulting
effective superpotential of the IR theory can be only \emph{approximately}
nongeneric, and metastability can still be preserved. (Note that we
emphasize {}``at leading order''; if those operators that destabilize
the metastable minimum are small enough, then the decay time of the
false vacuum is still sufficiently long to avoid the possibility of
decay within the lifetime of the Universe.) The nett effect can be
the lifting of the R-axion masses, and the radiative generation
of large gaugino masses. 

In ISS-type models, the breaking of both gauge symmetry and SUSY
at the metastable minimum, can be traced back to the gauge singlet field $\Phi_{i}^{j}$ ($i,j$
are ISS-sector flavour indices) of R-charge 2. As already stated,
we will require that R-symmetry is broken in the full theory, but
that this must be communicated to the low-energy effective theory in a controlled
manner so as to maintain the metastability -- as a working example
we will consider additional $SU(3)$ coloured fields with mass terms.
The resulting effective IR theory is non-generic, but this is naturally
understood as resulting from the dynamical breaking of R-symmetry.
At one-loop and higher this leads to R-symmetry breaking operators
of the form
\begin{equation}
W_{R}\approx\, {\rm const}\,\frac{Tr(\Phi)}{m_{R}}W^{\alpha}W_{\alpha},
\label{eq:loop}
\end{equation}
 where $W^{\alpha}$ is the gluon field-strength superfield, 
 $m_{R}$ represents the scale of R-symmetry breaking, and
${\rm const} \sim1/16\pi^{2}$ takes into account
the loop suppression (if this operator is generated perturbatively). In order to motivate $W_{R}$,
in the Appendix we will
show how \eqref{eq:loop} can be generated at one loop. It is important
that the R-symmetry violating terms \eqref{eq:loop} are holomorphic. Because
of this they have a relatively gentle effect on the vacuum structure
of our model; with or without them, SUSY is still broken in the metastable
minima, and is completely restored in the same SUSY preserving minima.
What these operators do is transmit the SUSY breaking from $\Phi$
to the gauginos. 

Indeed, in the metastable vacuum, SUSY is broken by $F_{\Phi_{i}^{i}}=h\mu^{2}$
(for some $i$) where $h$ and $\mu$ are parameters
of our choosing (in the low-energy magnetic theory), and the scale of $SU(2)_{L}$ 
breaking is $M_{W}\approx g_{2}\mu$
(note that for more general, i.e. flavour dependent, choices of $\mu$
this scale is an \emph{upper bound} on the scale of supersymmetry
breaking). The gaugino masses generated by these terms are then
\be
M_{\lambda}\, \approx\, {\rm const}\,\frac{h}{g_{2}^{2}}\frac{M_{W}^{2}}{m_{R}},
\ee
and the SUSY breaking induced in the gauginos $\lambda$ is in turn transmitted
to the squarks and sleptons through one-loop diagrams. 

This evades the first "no-go" theorem, 
provided that gaugino mass terms are large
enough, but how can one accommodate large $h$ in order to evade the second "no-go" theorem 
and overcome the loop suppression?%
\footnote{ Note that we are not entitled to take $m_{R}\ll1$ TeV because the
approximation in calculating $W_{R}$ breaks down.%
} Naively it seems impossible that the SUSY breaking in the visible
sector could ever be large enough: as in gauge-mediation scenarios
we require $F_{\Phi} \gtrsim 100$ TeV$^{2}$ in order to overcome the
one loop suppression factor $1/16\pi^{2}$, and the only
way to reconcile this with the fact that $M_{W}\approx g\mu$ is to
choose a coupling $h\gg1$. One might worry that this would render the theory
completely incalculable. We will argue that this is not the case,
and that the theory is calculable, at least at energies around the electroweak scale.
Ultimately
the effective theory is only slightly more complicated than the MSSM
itself, and yet the soft-SUSY breaking terms are essentially unsuppressed
and completely predicted. The spectrum and phenomenology is expected
to be broadly similar to gauge mediation \cite{GM} (see \cite{GM2,Kane} for
a review). One striking difference though is that the Higgs fields
become  very heavy and essentially decoupled from the theory.

Note that there is a connection here with Ref.~\cite{Dine,DM},
which provided a general framework for considering MSB models with R-symmetry
broken by small parameters. The present work can be thought of as
a concrete and minimal realisation of this idea, and indeed at very
low scales (i.e. scales far below $\Lambda_{L}$) our model is in
effect an example of a {}``retrofitted'' O'Raifeartaigh model. The
novelty here is that SUSY breaking and electroweak breaking
emerge from the same sector and that there are no hidden sectors.

We begin in the following section by reviewing ISS models and introducing
the proposed MSSM-like extension of it. As we have already indicated,
it is the $SU(2)_{L}$ factor of the gauge group which we are suggesting
plays the role of the magnetic dual, and which becomes strong at the
scale $\Lambda_{L} \gtrsim M_{Pl}$. However, introducing the other multiplets and
gauge groups necessary to make the model MSSM-like complicates the
vacuum structure. The dynamical restoration of SUSY which leads to
the supersymmetric minima involves various parameter-dependent combinations
of $F$ and $D$-flat directions; we will show explicitly how the
minima are generated and where they are located. We will also make
the following observation: the distance in field space from the metastable
origin to the global SUSY restoring minima can be less than, or greater
than the Landau pole $\Lambda_{L}$ and depends on the couplings.
Thus, by adjusting couplings, the SUSY restoring minima can be banished
beyond the Landau pole. 
Following this we discuss the generation of R-symmetry breaking terms, 
and then discuss the resulting phenomenology, and in particular 
its similarity to gauge mediation.

\section{MSSM: M is for metastable}

Can ISS metastability be successfully embedded in the visible sector
of the MSSM? In this section we demonstrate that with minor modification,
the MSSM can be transformed into a theory that has all the properties
of the macroscopic ISS theory, namely

\begin{enumerate}
\item O'Raifeartaigh SUSY breaking at the origin
\item A commensurately Higgsed electroweak sector
\item The gauge group which is Higgsed is IR free and has a Landau pole
at scale $\Lambda_{L}$
\item SUSY preserving global minima that are generated by strong dynamics
\end{enumerate}
This last point is enough to guarantee that the ISS-style metastability
has long decay lifetimes since the SUSY preserving minima are generated
radiatively and hence the potential is much flatter than it is broad
\cite{ISS}. In addition it guarantees that the metastable minima are
preferred in a Universe whose temperature is greater than the SUSY
breaking scale \cite{ACJK,AJK}. 

Note that there do exist chiral models with Seiberg duals, and even
chiral Seiberg duals of nonchiral theories, but as yet there are no
microscopic theories whose magnetic duals are MSSM-like; therefore
we will in what follows be working entirely in the IR macroscopic
theory and merely be assuming that a UV microscopic theory exists
above the Landau pole. This theory may or may not be a field theory
which may or may not be related to the macroscopic theory by Seiberg
duality, however this is irrelevant to our discussion. In particular
we only need show explicitly that the macroscopic theory inherits
the same O'Raifeartaigh-like metastable minima as in the explicit
ISS theory, and that issue is independent of the microscopic theory. 
Therefore let us first recapitulate the metastable SUSY breaking minima
of ISS.

\subsection{The ISS model}

Intriligator, Seiberg and Shih \cite{ISS} examined the IR free magnetic dual of an asymptotically
free $SU(N_{c})$ theory, in which the magnetic theory has a gauged
$SU(N_{f}-N_{c})$ symmetry and global $SU(N_{f})\times U(1)_{B}\times U(1)_{R}$
symmetry for degenerate quark mass terms in the microscopic theory.
The superpotential of the macroscopic theory is of the form
\be
W_{cl}=h[Tr(\varphi\Phi\tilde{\varphi})-\mu^{2}Tr(\Phi)]
\ee
where $\Phi_{j}^{i}$ are the flavour mesons of the IR free theory
and $\varphi_{i}^{a}$ and $\tilde{\phi}_{a}^{j}$ the fundamental
and antifundamentals of quarks under $SU(N_{c}-N_{f})$. 
 The crucial
observation is that for $N_{f}>N_{c}$ the F-flatness equation 
is no longer satisfied due to the
so-called rank condition; that is 
\be
F_{\Phi_{j}^{i}}\, =\, h\, (\tilde{\varphi}^{j}.\varphi_{i}-\mu^{2}\delta_{ij})\, =\, 0
\label{rank-cond}
\ee
can only be satisfied for a rank-($N_{f}-N_{c}$) submatrix of the
$F_{\Phi}$. The quark mass of the electric theory corresponds to
$\mu^{2}$ which is a free parameter. It could be generated in a variety
of ways \cite{ISS,DM,MN,AS}, but here we take it
as simply the control parameter for both SUSY and gauge breaking.
The model is of the standard O'Raifeartaigh type, with supersymmetry
being broken at scale $\mu$. The height of the potential at the metastable
minimum is given by 
\be
V_{T=0}(0)=N_{c}|h^{2}\mu^{4}|,
\ee
i.e. there is an equal contribution from each of the non-zero $F_{\Phi}$-terms.
The supersymmetric minima are located by allowing $\Phi$ to develop
a vev. The $\varphi$ and $\tilde{\varphi}$ fields
acquire masses of $\langle h\Phi\rangle$ and can be integrated out,
upon which one recovers a pure $SU(N_{f}-N_{c})$ Yang-Mills theory
with a nonperturbative contribution to the superpotential of the form
\be
W_{dyn}=N\left(\frac{h^{N_{f}}\mbox{det}_{N_{f}}\Phi}{\Lambda^{N_{f}-3N}}\right)^{\frac{1}{N}}.
\ee
This leads to $N_{c}$ nonperturbatively generated SUSY preserving
minima at 
\be
\langle h\Phi_{i}^{j}\rangle=\mu\epsilon^{-(\frac{3N_{c}-2N_{f}}{N_{c}})}\delta_{i}^{j}
\ee
in accord with the Witten Index theorem, where $\epsilon=\mu/\Lambda$.
The minima can be made far from the origin if $\epsilon$ is small
and $3N_{c}>2N_{f}$, the latter being the condition for the magnetic
theory to be IR-free. However since we must also have $N_{f}\geq N_{c}+2$
the positions of the minima are bounded by the Landau pole and they
are always in the region of validity of the macroscopic theory. 
We shall now adapt this structure to mimic the MSSM.

\subsection{A metastable MSSM}

The main content of the model will be a direct extension of
the supersymmetric Standard Model incorporating the above mechanism. 
Our goal is to embed the metastable ISS model into the Standard Model.
Since we are after a minimal such embedding we do not want to treat the ISS model
as a hidden sector. Instead we identify the $SU(N)$ group of the (magnetic) ISS theory 
with the $SU(2)_L$ weak-interaction gauge group of the Standard Model.
The $N_{f}$ pairs of  fields
$\varphi$ and $\tilde{\varphi}$ are then the Higgs doublets of the Standard Model.
(Of course now $N_{c}$
has no meaning other than as the combination $N_{f}-N=N_f-2$, nevertheless
we will retain it as a useful parameter.) The Higgs
sector superpotential of our model is taken to be
\be
W_{Higgs}=h\, Tr[\varphi\Phi\tilde{\varphi}-\mu^{2}\Phi],
\label{Higgs-sup}
\ee
so that electroweak symmmetry will be automatically Higgsed at the
same time as SUSY is broken. The parameter $(\mu^{2})_{i}^{j}=\mu_{i}^{2}\delta_{i}^{j}$
can be generated dynamically, and we will without loss of generality
take a flavour basis in which it is a diagonal but non-degenerate
matrix in flavour space, so that the $U(N_{f})$ flavour symmetry
is explicitly broken to $U(1)^{N_{f}}$. We now observe that in
order to have SUSY broken by the rank condition we must have $N_{f}\geq3$.
For simplicity we will concentrate here on the minimal case, $N_f=3$,
and note that our construction can be trivially extended to higher values of $N_f$. 
So we have to extend the MSSM to a multi-Higgs model with
three generations of Higgs pairs. (It will turn out that only one pair of Higgs doublets
will be coupled to the SM matter fields.) This is the first and last
modification that we need to make to the MSSM-sector; the remaining
fields $Q$, $U$, $D$, $L$, $E$ have the usual charge assignments
and number of generations.\footnote{ 
As we have already mentioned, 
we will also require an additional
sector which breaks the R-symmetry.
It is the coupling of this sector to the MSSM sector which yields
gaugino masses. However these fields will be chosen so they are a 
mild perturbation of the vacuum structure, metastability and 
spectrum of the MSSM-like 
sector of the theory, which is the subject of this section.}

We now want to discuss the symmetry of our model. When all $\mu_i$ parameters in \eqref{Higgs-sup}
are set to zero, this superpotential has $U(3)$ global flavour symmetry. We now turn on
the non-vanishing and non-degenerate values for $\mu$'s and order them so that 
$|\mu_1| > |\mu_2| > |\mu_3| > 0.$ This breaks the flavour symmetry down to
$U(1)^{3}$ which we denote as $U(1)_Y \times U(1)_{3} \times U(1)_{PQ}.$ We choose the charges
of $\p_i$ under these three $U(1)$'s as follows:
\be
U(1)_Y\, :\, (-\hf,\hf,0) \ , \qquad 
U(1)_3\, :\, (0,0,1) \ , \qquad 
U(1)_{PQ}\, :\, (1,1,1) 
\ee
The charges of $\pt_i$ fields are opposite, and the $\Phi_{ij}$ field transforms in the 
bifundamental representation under the first two $U(1)$'s such that the superpotential \eqref{Higgs-sup}
is invariant. We choose this assignment of charges (rather than two traceless combinations and one trace) 
for reasons that will become clear immediately. The PQ symmetry is the overall $U(1)$ of the broken $U(3)$
flavour symmetry\footnote{Note that the PQ symmetry was the {}``baryon number'' of the original
ISS model -- we do not use that name to avoid confusion with the conventional
$B$ of the MSSM. %
}.
We keep it as a global $U(1)$ of our model which is spontaneously broken by the vevs of $\p_1$ and $\p_2$ 
in the metastable vacuum. 
The $U(1)_Y$ symmetry gives rise to the hypercharge when quark and lepton superfields are included.
This symmetry is gauged and participates in the spontaneous electroweak symmetry breaking
$SU(2)_L \times U(1)_Y \, \to \, U(1)_{QED}$ by the vevs of $\p_1$ and $\p_2$.

In addition, there is also an (anomalous) R-symmetry $U(1)_R$ as well as the Baryon and Lepton number
symmetries.
We list the particle
content and charges of our metastable SUSY Standard Model
 in the Table%

\begin{center}\begin{tabular}{|c|c|c|c|c|c|c|c|c|c|}
\hline 
&
{\small $SU(2)_{L}$}&
$U(1)_{Y}$&
$\quad U(1)_3 \quad $&
{\small $\quad U(1)_{R}\quad $}&
$\quad U(1)_{PQ} \quad $&
{\small $\quad L \quad $}&
{\small $\quad B \quad $}\tabularnewline
\hline
\hline 
$\Phi_{ij}$&
$1$&
{\small $\frac{1}{2}(\delta_{i1}-\delta_{i2}+\delta_{j2}-\delta_{j1})$}&
{\small $\frac{1}{2}(\delta_{j3}-\delta_{i3})$}&
2&
0&
0&
0\tabularnewline
\hline 
{\small $\varphi_i$}&
$\square$&
{\small $-\frac{1}{2}\, , \, +\frac{1}{2}\, , \, 0$}&
{ $0\, ,\, 0\, , \, 1$}&
0&
$1$&
0&
0\tabularnewline
\hline 
{\small $\tilde{\varphi}_i$}&
$\bar{\square}$&
{\small $+\frac{1}{2}\, ,\, -\frac{1}{2}\, , \, 0$}&
{ $0\, , \, 0\, ,-1$}&
0&
$-1$&
0&
0\tabularnewline
\hline 
$L$&
$\bar{\square}$&
{\small $-\frac{1}{2}$}&
$0$&
1&
{\small$-\frac{1}{2}$}&
1&
0\tabularnewline
\hline 
$E$&
1&
$1$&
$0$&
1&
{\small $-\frac{1}{2}$}&
-1&
0\tabularnewline
\hline 
$Q$&
$\bar{\square}$&
{\small $\frac{1}{6}$}&
$0$&
1&
$-\frac{1}{2}$&
0&
{\small $+\frac{1}{3}$}\tabularnewline
\hline 
$D$&
1&
{\small $\frac{1}{3}$}&
$0$&
1&
{\small $-\frac{1}{2}$}&
0&
{\small $-\frac{1}{3}$}\tabularnewline
\hline 
$U$&
1&
{\small $-\frac{2}{3}$}&
$0$&
1&
{\small $-\frac{1}{2}$}&
0&
{\small $-\frac{1}{3}$}\tabularnewline
\hline
\end{tabular}\par\end{center}

The $U(1)_{R}$ symmetry will be
broken by the R-symmetry sector which we add later. As mentioned earlier, the hypercharge
of the Higgs fields is associated with the traceless $U(1)_Y$
factor of the parent flavour symmetry. The hypercharges of the remaining
fields are determined by anomaly cancellation under this $U(1)$.
These fields do not (and indeed need not) fall into obvious representations
of the parent flavour symmetry. The $U(1)_{Y}$ hypercharge factor
we shall assume to be gauged, and the 
factor of $U(1)_3$ 
is assumed to be global, and it will remain unbroken. The third Abelian factor
surviving from the broken flavour symmetry -- $U(1)_{PQ}$ -- is broken spontaneously by the vevs
of  $\p$ and $\pt$. This implies that there is a single Goldstone boson -- the PQ-axion --
present in our model in the metastable vacuum (after the three Goldstone bosons of the $SU(2)_L \times U(1)_Y$
gauge group are eaten by the longitudinally polarized vector bosons). It follows that 
apart from the PQ axion which we will discuss in a moment, there will be no
massless scalars arising from the metastable vacuum of our model.

The symmetries allow the masses of the quarks and leptons
to be generated by the standard Yukawa couplings of the MSSM, 
\be
W_{Yuk}=\lambda_{U}Q\varphi_{2}U+\lambda_{D}Q\varphi_{1}D+\lambda_{E}L\varphi_{1}E,
\label{W-yuk}
\ee
where the $\lambda_{f}$ carry conventional MSSM generation indices.
Note that with our assignment of charges, and in particular the 
$U(1)_{PQ}$ and $U(1)_3$ symmetry, these are all the Yukawa couplings one can write down.
The superfields $\varphi_1$ and $\varphi_2$ are the two Higgs doublets,
$H_d$ and $H_u$ of the MSSM, and all the remaining ISS chiral fields (i.e. 
$\p_3$, $\pt_1$, $\pt_2$, $\pt_3$ and $\Phi_{ij}$) 
cannot couple to the quarks or leptons. This avoids flavour changing
neutral currents appearing at tree-level. 

The $U(1)_R$ symmetry in our model is broken by anomalies to a discrete symmetry which contains $Z_2$.
This $Z_2$ is the conventional R-parity which protects against 
baryon and lepton number violating operators.

Ultimately the only light state remaining in the Higgs sector will be
the axion, whose mass is protected by the anomalous PQ symmetry.
As it stands this (visible) axion would be disallowed because the
Peccei-Quinn scale would be ${\cal{O}}(M_W)$. There are two natural
ways to make such an axion  acceptable:
either the PQ symmetry breaking scale is elevated to $f_{PQ}\sim 10^{11}$ Gev
by some other hidden sector fields which are also charged under $U(1)_{PQ}$
in the sense of \cite{DF},
or the PQ symmetry is gauged. 

For the purposes of discussion in this paper
we will assume the latter.  In this case we add an additional field $\eta$ to the microscopic
theory, which is charged under the $U(1)_{PQ}$ and gets a very large vev $\gg \mu.$  
The PQ-axion is the Goldstone boson which is eaten by the longitudinal mode of the
massive gauge boson and effectively acquires the mass $g \langle \eta \rangle \gg M_W.$
Thus it disappears from the light spectrum.
The anomaly is generically cancelled by the Green-Schwarz mechanism in string theory.
This generally (although not always) induces Fayet-Iliopoulos (FI) terms
into the $D$-terms of the Lagrangian. This gives a
natural mechanism for generating the vev $\langle \eta \rangle$ 
if this is the field
which cancels the FI term.

An alternative approach, which we have not explored in this paper, but
will study elsewhere, is to allow the eventual R-symmetry breaking to also
induce PQ-symmetry breaking mass terms for the Higgs fields of the
form $W_{Higgs-mass}\sim m_{ij} \varepsilon_{ab} \varphi^a_i \varphi^b_j $.
These terms would be equivalent to the "$\mu$"-terms of the
conventional MSSM and would break both $PQ$ and R-symmetry.
We will leave a full discussion of these issues to future work.

\subsection{The local metastable minima}

Let us identify the metastable supersymmetry breaking vacuum of our model. The analysis 
here follows Ref.~\cite{ISS} closely, but with the important modification that 
flavour symmetry is broken by the $\mu_i$ terms, so that 
there are no (uneaten) Goldstone modes.
The rank condition ensures that SUSY is broken at the origin of $\Phi$,
and $SU(2)_{L}$ is Higgsed. Without loss of generality take 
$|\mu_{1}|>|\mu_{2}|>|\mu_{3}| > 0$.
The $D$-terms in the potential ensure that $\varphi^{i}=\tilde{\varphi}_{i}$,
and the $F$-terms for the Higgs sector are of the form 
\be
V_{F}=|h|^{2}|\varphi^{i}\tilde{\varphi}_{j}-\mu^{2}\delta_{j}^{i}|^{2}+
|h|^{2}|\Phi_{i}^{j}\tilde{\varphi}_{j}|^{2}+
|h|^{2} |\varphi^{i}\Phi_{i}^{j}|^{2},
\label{Higgs-VF}
\ee
so that, as in the original ISS model, the Higgs vevs are at 
\be
\varphi=\tilde{\varphi}^{T}=\left(\begin{array}{ccc}
\mu_{1} & 0 & 0 \\
0 & \mu_{2} & 0 \end{array}\right),
\ee
the non-zero $F$-term is  $F_{\Phi_{3}^{3}}=h\mu_{3}^{2}$,
and the height of the potential at the metastable minimum is given
by 
\be
V(0)=|h^{2}\mu_{3}^{4}|.
\ee
The vevs can be written succinctly using a block notation
\be
\Phi=\left(\begin{array}{cc}
Y & Z\\
\tilde{Z} & X\end{array}\right)\,\,;\,\,\,\varphi^{T}=\left(\begin{array}{c}
\sigma\\
\rho\end{array}\right)\,\,;\,\,\,\tilde{\varphi}=\left(\begin{array}{c}
\tilde{\sigma}\\
\tilde{\rho}\end{array}\right)
\ee
 where for example $Y$ is a $2\times2$ matrix and $X$ is 
(for the minimal assumption that $N_{f}=3$) a single field.
Defining flavour indices $\ui=1,2$ and colour indices $a=1,2$ the vevs
are 
\begin{eqnarray*}
\langle\sigma \rangle & = & \langle\tilde{\sigma}\rangle=\left(\begin{array}{cc}
\mu_{1} & 0\\
0 & \mu_{2}\end{array}\right)\\
\langle\rho_a \rangle & = & \langle\tilde{\rho}_a \rangle \, = \, 
\langle Y\rangle_{\ui}^{\uj} \, = \,  
\langle Z\rangle_{\ui} \, = \, 
\langle\tilde{Z}\rangle^\ui \, = \, 0 \\
\langle X\rangle & = & X_{0}\, .\end{eqnarray*}
It is straightforward to identify those fields that gain a mass at tree level
from the $F$-terms. Defining the eigenstates 
(for simplicity we take $\mu_i^2$ to be real in the
remainder of this subsection)
\be 
\rho^a_\pm = \frac{1}{\sqrt{2}}
(\rho\pm \tilde{\rho}^*)^a 
\,\,\,\, ; \,\,\,\,
(\sigma_\pm)^a_\ui = \frac{1}{\sqrt{2}}
(\delta\sigma\pm \delta\tilde{\sigma})^a_\ui
\ee
we find the $F$-term contributions to the mass-squareds {\it at tree level} to be
\begin{eqnarray}
m_{\rho^a_+}^2 & = & h^2 (\mu_a^2+\mu_3^2 ) \nonumber \\
m_{\rho^a_-}^2 & = & h^2 (\mu_a^2-\mu_3^2 ) \nonumber \\
m_{\sigma_+}^2 & = & h^2 \mu_\ui \mu_a \nonumber \\
m_{Y_{\uj}^\ui}^2 & = & h^2 (\mu_\ui^2+
\mu_{\uj}^2 ) \nonumber \\
m_{Z_\ui}^2 & = & m_{\tilde{Z}^\ui}^2 \,  = \, 
h^2 \mu_{\ui}^2 \nonumber \\
m_{\sigma_-}^2 & = & m_X^2 \, = \, 0 
\, . 
\end{eqnarray}
Note that the higher minima (i.e. those involving $F_{\Phi_{1}^{1}}=h\mu_{1}^{2}$
or $F_{\Phi_{2}^{2}}=h\mu_{2}^{2}$ instead of $F_{\Phi_{3}^{3}}\neq 0$) are unstable to decay into this
one since the $\rho_-$ and $\tilde{\rho}_-$ become tachyonic. Also in the 
case of degenerate $\mu_i$ these same states become additional Goldstone modes
reflecting the enhanced flavour symmetry. The traceless part of 
the states $Re(\sigma_-)$ are lifted by the $D$-terms, and the 3 traceless 
components of $Im(\sigma_-)$ are eaten to become the longitudinal components of the 
$W^\pm$ and $Z$. This leaves $X$ and $Tr(\sigma_-)$ as pseudo-moduli, 
the latter being associated with the spontaneously broken but anomalous 
PQ symmetry. As in \cite{ISS} one can now evaluate the 
one-loop contribution to these mass-squared and find that they are  
${\cal{O}}(h^4 \mu^2/16\pi^2 )$. 

We conclude that all these Higgs fields are massive 
and in the limit $h \gg 1$ become very heavy. In this large-$h$ limit
all Higgs fields will decouple from physics at and around the electroweak scale.
The PQ-axion is also removed from the spectrum by choosing the unitary gauge 
of the gauged PQ-symmetry as we have explained above.

To complete the discussion of the model around the metastable vacua,
let us estimate the position of the Landau pole. This can be evaluated
from the usual expression for the Wilsonian gauge coupling with
a beta-function which, for the particle content listed, is $b_0=-N_f$:
\be
e^{-\frac{8\pi^2}{g_2(\mu)}}\, =\, \left( \frac{\mu}{\Lambda_L} \right)^{N_f} \, .
\ee
Taking $N_f=3$, $\alpha_{SU(2)} = 1/30 $ and $\mu=100$~GeV the Landau
pole is found to be comfortably much greater than the Planck scale.
Since at the Planck scale the physics is supposed to be modified anyway by 
the inclusion
of gravity, knowledge of the electric theory is not even required.
If extra $SU(2)$ fields are introduced into  the model or into the
R-messenger sector, the number of flavours may be
increased (as we shall see later). 
Two or more extra flavours implies that there
is a $\Lambda_L$ below $M_{Pl}$. In what follows we will always assume that the UV completion
of the theory would be required at $\Lambda_L \sim M_{Pl}.$

\subsection{The global minima: dynamical restoration of SUSY}

Having established the existence of non-supersymmetric vacua,
now we want to show that in this model there are also supersymmetric ground states.
In these `true' vacua supersymmetry will be restored dynamically as in the ISS model.
To find these vacua we will first need to determine the dynamical superpotential of the model
and then solve the resulting F-flatness equations.

The behaviour of the $SU(2)_{L}$ group factor will of course be affected
by the extra doublets of the Standard Model. The number of extra $SU(2)_{L}$ fundamentals
is $12$ (i.e. three generations each of $L$, and of $Q\times 3$ colours); in order
to be general we shall call this number $n,$ and also will use $SU(N)$ for $SU(2)_L$,
\be
3(L+ 3\times Q) = 12\, :=\, n \quad , \qquad SU(2)_L\,:=\,SU(N).
\ee
The first coefficient of the $\beta$-function of the $SU(N)$ gauge theory is
$b=3N - \hf n -N_f.$
As our first step,
we note that dynamical supersymmetry restoration requires vevs
along more than just the $\Phi$ direction. Indeed giving large vevs
to $\Phi$ gives masses to $2N_{f}$ fundamental fields ($\p_i$ and $\pt_i$), so the beta
function at scales below these masses becomes $b=3N-\hf n$. Since we wish to have $N=2$ and $n=12$
this gives a $\beta$-function for $SU(2)_{L}$ which is coincidentally
zero and we conclude that there can be no dynamically generated term
which is solely dependent on $\Phi$. Clearly dynamical supersymmetry
breaking requires that we integrate out more flavours to reverse the
sign of the $\beta$ function. In order to do this we must search
along directions that give masses to the other fundamentals of $SU(N)$,
but do not break $SU(N)$ and that are gauge invariant monomials (i.e.
$D$-flat). As well as $\Phi$ itself, there are 42 independent monomials
that one could consider
\be
UDD\,\,\,;\,\,\, UUUEE\,\,\,;\,\,\, UUDE,
\ee
where we have suppressed flavour indices \cite{Gherghetta}. 
In general, giving vevs to a combination of these directions
will give masses to some of the flavours which can then be integrated
out. 

In our model all fundamental matter fields ($n$ of $Q$'s
and $L$'s and $2N_f$ of $\p$'s and $\pt$'s) will become massive along these directions and
can be integrated out. In the IR the theory will become a pure SYM with the gauge group $SU(N)$.
This gauge group confines, and the expression for the
nonperturbative superpotential $W_{dyn}$
of this IR theory is uniquely determined by the gaugino condensation to be
\be 
W_{dyn}\,=\, N (\Lambda_{SU(N)})^3.
\label{low-e-sup}
\ee
Here $\Lambda_{SU(N)}$ is the dynamical scale of the pure SYM.
To derive the dynamical superpotential of the original theory 
with all matter fields present we use \eqref{low-e-sup} and the matching
relations for the $\Lambda$ scales of the theory below and above each mass threshold.
In order to keep the discussion general, let us assume 
that  the $i$'th set of fundamentals is integrated out
at the scale $E_{i}$, whereupon the $\beta$ function changes from
$b_{i}\rightarrow b_{i+1}$. Specifically we take $i=0$ at the highest scale so that
\be
b_0 = b = 3N - \hf n -N_f \quad , \qquad \Lambda_0 = \Lambda_L
\ee
and take $E_1 \ge E_2 \ge \ldots \ge E_i \ge E_{i+1} \ge \ldots E_n$
to be the masses of each of the $n$ matter fundamentals $Q$ and $L$.
The matching of Wilsonian gauge coupling constants at each mass threshold $E_i$
gives
\be
e^{-8\pi^{2}/g^{2}(E_{i})}=(E_{i}/\Lambda_{i})^{-b_{i}} \, =\,
(E_{i}/\Lambda_{i-1})^{-b_{i-1}} \quad => \quad
\Lambda_{i}^{b_{i}}\,=\, \Lambda_{i-1}^{b_{i-1}}  E_i^{b_{i}-b_{i-1}}
\ee
These relations relate $\Lambda_0=\Lambda_L$ of the high-energy theory to the
scale $\Lambda_n$ of the gauge theory with all $n$ fundamentals integrated out. We now need to
integrate out the remaining $2N_f$ of $\p$ and $\pt$ matter fields to descend to the pure SYM with
$\Lambda_{SU(N)}$. Masses of $\p$ and $\pt$ fields are set by the vevs of $\Phi_{ij}$
so that the mass to the $N_f$ power is given by
\be 
\left(m_{\p,\pt}\right)^{N_f} \, =\, h^{N_f} \, \det_{N_f} \langle \Phi \rangle
\ee

We thus obtain the following general expression for the dynamically generated superpotential 
\be
W_{dyn}\, =\, N\left(\Lambda_L^{3N-N_{f}-n/2}h^{N_{f}}(\mbox{det}_{N_{f}}\Phi)\,
\prod_{i=1}^{n}E_{i}^{(b_{i}-b_{i-1})}\right)^{\frac{1}{N}}
\ee
where $b_{i}=3N-N_{f}-\hf (n-i)$. Note that
the dependence on any particular energy scale is dependent only on
the change in $\beta$-function from the states that are integrated
out there, so that $W_{dyn}$ behaves correctly if we let any of the thresholds
coalesce. As long as the eventual $\beta$-function is positive (i.e. $b_0$ is negative) at
the high scale we are assured of generating such a dynamical superpotential, $W_{dyn}$.

We can 
now solve the F-flatness condition for the superpotential,
\be
\frac{\partial}{\partial \Phi} \left( W_{dyn} -h (\mu_i \Phi_{i}^i)\right) \, =\, 0
\ee
to find
\be
h\Phi_{kk}\,=\,
\frac{1}{\mu_k^2} \left[ \prod_{i=1}^{N_f}\mu_i^2 \right]^{1/(N_f-N)}
\left(\Lambda_{L}^{3N-N_{f}-n/2}\,
\prod_{i=1}^{n}E_{i}^{(b_{i}-b_{i-1})}\right)^{-\frac{1}{N_f-N}} \quad , \quad k=1,\ldots,N_f
\label{Phimin}
\ee
 Substituting back into $W$ we find
\be
W_{min}\,=\, 
N_{c} \left[ \prod_{i=1}^{N_f}\mu_i^2 \right]^{\frac{1}{N_{c}}}
\left(\Lambda_{L}^{3N-N_{f}-n/2}\,
\prod_{i=1}^{n}E_{i}^{(b_{i}-b_{i-1})}\right)^{-\frac{1}{N_{c}}} \quad , \quad N_c:=N_f-N
\ee
As in the original ISS model we appear to be running to $N_{c}$ minima
in $\Phi$. Here $N_c$ does not necessarily have an intrinsic meaning beyond its definition $N_c=N_f-N$,
but we will use it in formulae below. 

We now replace the thresholds $E_{i}$ with holomorphic fields corresponding
to the masses induced by non-zero vevs of gauge invariant monomials.
We shall consider an R-parity conserving theory in order to avoid
proton-decay, in which case the dynamically generated superpotential
can only depend on the 27 $UUDE$ monomials. Turning on a vev
in this direction, which we shall refer to as $X_{1}^{4}$,
\be
X_1^4 \,=\, UUDE
\ee
 generates
masses for the quark and lepton fields via the standard
Yukawa couplings of the MSSM. 
These masses are given by the
$|F_{\varphi}|^{2}$ and $|F_{\tilde{\varphi}}|^{2}$ terms in the
potential, and are 
\begin{eqnarray}
E_{Q} & \equiv & \sqrt{2|\lambda_{U}^{2}|+|\lambda_{D}^{2}|}\, X_{1}\\
E_{L} & \equiv & \lambda_{E}X_{1}
\end{eqnarray}

With this prescription the potential is a runaway to large values
of $X_{1}$. However this is because as it stands the potential is
\emph{nongeneric} (in the sense of Nelson and Seiberg \cite{NS}). Consider
making the superpotential \emph{generic} by adding to it the nonrenormalizable
term $\frac{\lambda}{M}UUDE$ where $M \gtrsim \Lambda_{L}$ is a high mass scale,
which is allowed by all the symmetries of the theory including R-parity
(note that the fact that $X_{1}$ appears in $W_{dyn}$ means that
this term \emph{must} have been allowed in the original superpotential).
In order to satisfy constraints from proton decay, we shall implicitly
assume that $M\sim M_{Pl}$. 

To simplify the discussion let us integrate out all $n$ of the $SU(N)$
fundamentals at the scale $\lambda_{t}X_{1}$
(in other words, for simplicity of presentation we will not distinguish between the different
Yukawa couplings and we will also set all $\mu_i=\mu$)
\be
W_{min}=N_{c}\mu^{\frac{2N_{f}}{N_{c}}}\left(\Lambda_{L}^{3N-N_{f}-n/2}\,
(\lambda_{t}X_{1})^{n/2}\right)^{-\frac{1}{N_{c}}}+\frac{\lambda}{M}X_{1}^{4}
\ee
Minimizing this potential we find the supersymmetric minima at
\be
\langle X_{1}\rangle\,=\, \left(\left(\frac{nM}{8\lambda_{t}}\right)^{N_{c}}\,
\Lambda_{L}^{n/2+3N_{c}-2N_{f}}\mu^{2N_{f}}\lambda_{t}^{-n/2}\right)^{\frac{1}{4N_{c}+n/2}}.
\ee
To be inside the Landau-pole circle we would like $X_1$ to be less than $\Lambda_L$; we may rewrite
it as 
\be
\langle X_{1}\rangle\, =\, \Lambda_{L}\left(\left(\frac{n}{8\lambda_{t}}\right)^{N_{c}}\lambda_{t}^{-n/2}\,
\right)^{\frac{1}{4N_{c}+n/2}}\, \epsilon^{\frac{2N_{f}}{4N_{c}+n/2}} \, \xi^{\frac{-N_{c}}{4N_{c}+n/2}}
\ee
where
\textbf{
\begin{eqnarray}
\epsilon & = & \mu/\Lambda_{L}\\
\xi & = & \Lambda_{L}/M.
\end{eqnarray}
}
Setting all the Yukawas to be of order unity, the constraint $\langle X_{1}\rangle<\Lambda_{L}$
translates into $\xi > \epsilon^{\frac{2N_{f}}{N_{c}}}$.
This is easy to achieve as $\epsilon$ is naturally $\ll 1$ and $\xi$ can be $\lesssim 1$.
The second requirement is that $\mu \ll \langle \Phi \rangle < \Lambda_L$ where the first
inequality is needed to ensure that $\langle \Phi \rangle$ is far away from the origin and there
is no tunnelling to the metastable vacuum. From \eqref{Phimin} we have 
\be
\langle \Phi\rangle\,\sim\, h^{-1}
\mu^{\frac{2N}{N_c}}
\left(\Lambda_{L}^{3N-N_{f}-n/2}\,
\,\langle X_{1}\rangle^{n/2}\right)^{-\frac{1}{N_c}} 
\label{Phimin2}
\ee
In our model $N=2$, $N_f=3$, $N_c=1$ and $n=12$ and we have
\begin{eqnarray}
\langle \Phi\rangle &\sim& \Lambda_{L}\, h^{-1} \left(\frac{\mu}{\Lambda_L}\right)^{0.4}
\left(\frac{\Lambda_L}{M}\right)^{0.6} \ll \Lambda_L \\
\langle \Phi\rangle &\sim& \mu\, h^{-1} \left(\frac{\Lambda_L}{\mu}\right)^{0.6}
\left(\frac{\Lambda_L}{M}\right)^{0.6} \gg \mu
\end{eqnarray}
and both inequalities can be easily satisfied  for not too large $h$.
At very large values of $h$ the last equation, however, may not be valid and we will
discuss this case in the next subsection.

For now we conclude that
any generic superpotential (i.e. one which includes nonrenormalizable
operators allowed by the symmetries), dynamically restores the supersymmetry
at a scale below the Landau pole.

Of course one may worry that the nonrenormalizable operator is generated
by physics at a scale $M  \gtrsim \Lambda_{L}$ which is the region outside of validity
of the macroscopic theory. However this is perfectly consistent; the
microscopic theory is expected to generate all operators allowed by
the symmetries, and even if they are nonrenormalizable, they can be
simply rewritten in terms of the fundamental fields of the macroscopic
theory.

\subsection{(Lack of) tunnelling out of the metastable vacuum when $h\gg1$}

One question that we should clarify is metastability at $h\gg1$. As
has been argued in the literature, for $h\sim1$ it is rather easy
to find lifetimes for the metastable vacua that are longer than the
age of the Universe. The criterion is that there should be less than
one tunnelling event in the past light cone of the Universe which
translates into 
\be
S_{E}\gtrsim400
\ee
where $S_{E}$ is the Euclidean bounce action. As a rule-of-thumb,
the latter is given by 
\be
S_{E}\sim2\pi^2 \frac{\langle \Phi \rangle^{4}}{\Delta V},
\ee
so that potentials that are wider than they are tall have longer lifetimes.
In the pure ISS model this gives a not very severe bound on $\langle \Phi \rangle$.
In detail one finds that 
\be
S_{E}=\frac{2\pi^{2}}{3h^{2}}\frac{N^{3}}{N_{f}^{2}}\frac{\langle \Phi \rangle^{4}}{\mu^{4}},
\ee
which translates into 
\be
\frac{\langle \Phi \rangle}{\mu}\gtrsim3\sqrt{h}.
\ee
However as we have seen $\langle \Phi \rangle$ scales as $\mu h^{-1}$ times
by some large dimensionless functions. In pure ISS we find
\be
\frac{2\pi^{2}}{3h^{6}}\epsilon^{-4\frac{(3N-N_{f})}{(N_{f}-N)}}\gg400.
\ee
where $\epsilon=\mu/\Lambda_{L}$. In fact the power of $h^{-6}$ can
be found by a simple scaling argument without even evaluating the
action; defining $\hat{\Phi}=h\Phi$ and $\hat{x}=h^{2}x$ in the
bounce action one finds $S_{E}(h)=h^{-6}S_{E}(1).$ Now for certain
values of $N$ and $N_{f}$ this can make the bounds far more restrictive
for large $h$. If for example $N_{f}=5N/2$ then one requires $\epsilon\ll0.04\, h^{-\frac{9}{2}}$
which can be a severe bound on $\Lambda_{L}$. 

We now find a pleasant surprise for our model. As we observed, 
SUSY restoration involves
other flat directions of the MSSM. Although the supersymmetric minima
are indeed found at small values of $\langle \Phi \rangle$ as $h\gg1$, the vevs
along the other flat directions are independent of $h$, since they
depend only on the thresholds induced by the Yukawa couplings of the
MSSM-sector. Hence at large $h$ the rule-of-thumb still applies but
with $\langle \Phi \rangle$ replaced by the vevs of the other flat directions
which as we have seen can be as large as $\Lambda_{L}$. Thus large $h$
destroys the metastability in the pure ISS model, but in our model
it does not.

\section{R-symmetry breaking}

In order to generate Majorana masses for gauginos we need to break the R-symmetry
of the model. The easiest way to do this is by adding an appropriate higher-dimensional operator
to the superpotential. The unique leading-order operator of this type is of the form
\begin{equation}
W_{R}\, =\, \frac{g^2_{A}}{16\pi^2} \,\frac{Tr(\Phi)}{m_{R}}\,W_{A}^{\alpha}W_{\alpha}^{A},
\label{eq:loop2}
\end{equation}
where $W_A$ is the field-strength chiral superfield of the gauge field of type $A$ (where $A=1,2,3$ distinguishes
between the different types of gauge fields, $SU(3)$ colour, $U(1)_Y$ or $SU(2)_L$, in the Standard Model);
the factor of $g_A^2/(16\pi^2)$ comes from the fact that this operator is generated radiatively, and $m_R$ is the
mass scale of the R-symmetry violating effects.
In writing down \eqref{eq:loop2} we have made use of the fact that the $Tr(\Phi)$ is a gauge singlet field
and as such can be coupled to the combination $W_{A}^{\alpha}W_{\alpha}^{A}$.

This operator breaks R-symmetry since $\Phi$ has R-charge 2, and each of the $W$ fields has 
R-charge equal to 1. It can be straightforwardly generated by additional massive fields 
which couple to the gauge fields, which can either be integrated out or included as 
new degrees of freedom in the low energy theory. An example of the latter is discussed in the Appendix, 
where it is shown how the operator \eqref{eq:loop2} can be generated at one-loop. 
In the remainder of the paper we will 
simply assume the presence of the 
R-symmetry violating operator $W_R$. 

\subsection{Generation of gaugino masses}

The superpotential \eqref{eq:loop2} generates Majorana masses $M_{\lambda_A}$  for the gauginos $\lambda_A$ via
\be
\int d^2\theta\,\frac{g^2_{A}}{16\pi^2} \,\frac{Tr(\Phi)}{m_{R}}\,W_{A}^{\alpha}W_{\alpha}^{A} \, \ni\,
\frac{g^2_{A}}{16\pi^2} \,\frac{Tr\langle F_\Phi\rangle}{m_{R}}\,\lambda_{A} \lambda_{A}.
\ee
The vev for the $F$-term for $\Phi$ is non-vanishing in the metastable vacuum and is given by
\be
\langle F_{\Phi_{33}}\rangle  \,=\, -h \mu_3^2,
\label{F-Phi33}
\ee
which determines
the gaugino masses to be,
\be
M_{\lambda_A}\,=\, h\,\frac{g^2_{A}}{16\pi^2}\, \frac{\mu_3^2}{m_R}. 
\label{gaugino-mass}
\ee

It is remarkable that we are able to obtain these SUSY-breaking gaugino masses \eqref{gaugino-mass}
from the manifestly supersymmetric superpotential \eqref{eq:loop2}, \eqref{Higgs-sup}
directly and without adding any new degrees of freedom to our model. The gaugino masses are generated
by $\Tr \langle F_{\Phi}\rangle.$ The fact that $\Tr \langle F_{\Phi}\rangle \neq 0$ is of course the
consequence of the rank condition and is the key feature of our metastable model which is responsible for
SUSY breaking.

To get a rough estimate of the values of gaugino masses we can take $m_R \sim \mu_3 \sim 100 {\rm GeV}$ and
$h \sim 16\pi^2/g^2 \gg 1$. Then $M_{\lambda} \sim \mu \sim 100 {\rm GeV}.$
As we have already anticipated in the Introduction, we see that in order to get sizable gluino masses
we have to assume that the Higgs sector of our model is sufficiently strongly coupled, i.e. $h \sim 16\pi^2/g^2 \gg 1$.
In the following subsection we will 
argue  that this requirement of $h \gg 1$ does not render the theory incalculable,
instead it actually simplifies it at energy scales not much above the electroweak scale (i.e. below the
Higgs mass scale $M_H \sim h \mu \gg M_W \sim g\mu$).

\subsection{Theory at large h and decoupling of Higgses}

Clearly the perturbation theory in powers of $h$ breaks down when $h \gg 1.$
Here we would like to point out that this fact does not necessarily render our model incalculable.
Instead it signals the decoupling of all the Higgs fields from physics at scales low compared to their masses, 
i.e. at the electroweak scale.

The primary effect of large $h$ is to ensure that all the Higgs masses are very large.
We have already seen that treated in perturbation theory these masses receive 
their first non-vanishing contributions at the
order $h \mu$ or $\frac{h^2}{16\pi^2} \mu$. 
There of course can be significant corrections to these masses from higher orders of perturbation theory
in $h$. We will assume that the full masses of all the Higgses stay 
large, i.e. of the order $M_{Higgs} \sim h \mu$ or higher as $h \gg 1.$ 

When estimating quantum corrections from these heavy Higgs fields on internal lines to Feynman diagrams
for various processes at energies below $M_{Higgs}$ one would be required to use the full (resummed) masses.
It is then easy to see by power counting that the whole Higgs sector decouples from the Standard Model physics
at these energy scales. The $h^2$ enhancement of the Higgs self-interactions as in \eqref{Higgs-VF}
will be overcome by the factors of $h^2$ from the Higgs masses in the propagators. 

\section{Discussion}

We have presented an extremely compact formulation of the visible sector supersymmetry
breaking in a supersymmetric version of the Standard Model. 
Supersymmetry breaking is a consequence of the ISS-type metastable vacuum in our model
guaranteed by the superpotential 
\be
W_{Higgs}=h\, Tr[\varphi\Phi\tilde{\varphi}-\mu^{2}\Phi].
\label{Higgs-sup2}
\ee
The Yukawa interactions allowed by the symmetries of the model 
are precisely of the minimal standard form
\be
W_{Yuk}=\lambda_{U}Q\varphi_{2}U+\lambda_{D}Q\varphi_{1}D+\lambda_{E}L\varphi_{1}E.
\label{W-yuk2}
\ee
The SUSY breaking induced by the rank condition of \eqref{Higgs-sup2} 
is communicated to gauginos via the R-symmetry breaking
effective superpotential term
\begin{equation}
W_{R}\, =\, \frac{g^2_{A}}{16\pi^2} \,\frac{Tr(\Phi)}{m_{R}}\,W_{A}^{\alpha}W_{\alpha}^{A}.
\label{eq:loop32}
\end{equation}
In our model both supersymmetry and the electroweak gauge symmetry is broken by the same 
parameters $\mu_i \sim 100$ GeV. More precisely, vector bosons get their masses from $\mu_1$ and $\mu_2$
while SUSY is broken by $\mu_3$ where $|\mu_3| < |\mu_{1,2}|.$ 
This provides us with a non-trivial direct link between
SUSY breaking and the electroweak breaking. The price we have to pay for this very low SUSY breaking scale is
that we have to have strong coupling in the Higgs sector, $h \gg 1,$ in order to generate sufficiently large
gaugino masses from \eqref{eq:loop32}, \eqref{Higgs-sup2}.

However, we have argued that the primary role of large-$h$ is to make all the Higgs fields 
very heavy compared to the electroweak scale. At such energy scales all the Higgs fields become non-dynamical,
frozen at their vevs, and essentially the whole Higgs sector decouples from the theory. 

The masses for
squarks and sleptons in our model are generated from the gaugino masses above at one-loop level
in a way which is completely analogous to the gauge mediation scenarios \cite{Weinberg}, \cite{GM,GM2}.

In conclusion, we have constructed an extremely compact model of spontaneous supersymmetry breaking 
in the visible sector which gives a softly broken supersymmetric Standard Model as the
low-energy effective theory.

\section*{Acknowledgements}   

We thank Sakis Dedes, Ken Intriligator, Joerg Jaeckel and Stefan Forste for useful conversations.

  \renewcommand{\theequation}{A-\arabic{equation}}
  \setcounter{equation}{0}  
  \section*{APPENDIX:\,\,\, R-symmetry breaking with new massive fields}  

In the case that R-symmetry is broken by fields which remain in
the low energy theory, it may be necessary to add additional flavours (i.e. make $N_f=4$)
in order to enhance the rank condition and preserve metastability.
As an example, consider the following addition to the superpotential
(as in Ref.~\cite{MN}):
\be
 \kappa\,Tr(\Phi)\, \tilde{f}\cdot f \,+\, m_f \, \tilde{f}\cdot f \, ,
\label{f-sup}
\ee
where $f$ and $\tilde{f}$ are the new R-neutral fields transforming in the fundamental 
and anti-fundamental representations of the gauge group of type $A$,
and $\kappa$ and $m_f$ are constants.
In \eqref{f-sup} they are coupled to the gauge singlet $Tr(\Phi)$ and they have 
an R-symmetry violating mass-term $ m_f$. It is easy to see that at one-loop the 
$f$ and $\tilde{f}$ fields indeed give contributions of the type \eqref{eq:loop2}. 
They remain in the spectrum as massive fields coupled
to the gauge supermultiplets of type $A$, e.g. to gluons and gluinos. 

Adding $f$,$\tilde{f}$ to the theory is the simplest way of generating operators of the form 
eq.(\ref{eq:loop2}), but one should worry about the fact that 
the new R-symmetry breaking fields $f$ and $\tilde{f}$
in \eqref{f-sup} interfere with the rank condition \eqref{rank-cond}, possibly 
destroying the metastability of the non-supersymmetric vacuum. 
In the minimal model with $N_f=3$, 
the $F$-term $\langle F_{\Phi_{33}}\rangle $ can in principle be set to zero 
(and invalidate the rank condition) by turning on
large vevs $\langle f \rangle =\langle \tilde{f} \rangle =
f_0= \sqrt{h/\kappa}\, \mu_3,$ corresponding to a new supersymmetric vacuum 
a distance $f_0$ along the $f$-direction
from the metastable vacuum at the origin. This breakdown of the
rank condition is easily avoided 
by the simple expedient of adding a fourth generation of 
$\varphi$, $\tilde{\varphi}$. Note that this does not 
introduce any new Goldstone modes, since the new global $U(1)$ symmetry 
is exact and unbroken (just like the $U(1)_3$ before), and neither 
does it introduce new Yukawa couplings with the SM matter fields 
since the new $\varphi_4$,~$\tilde{\varphi}_4$ fields are neutral 
under hypercharge. 

It easily follows in $N_f= 4$ models, that
there are no (perturbative) supersymmetric vacua 
and ISS-type metastability is completely restored. In fact 
in this case there are two candidates for the metastable 
vacua. The first is similar to the one we have already explored, 
and is characterized by 
\begin{eqnarray}
\langle f.\tilde{f}\rangle & = & 0  \nonumber \\
\langle \Phi_{ij}\rangle & = & 0 \nonumber \\
\label{first}
\langle \varphi_{i}\tilde{\varphi}_{j}\rangle & = & \left(\begin{array}{cccc}
\mu_{1}^{2} & 0 & \,\,\, 0 \,\,\,  & \,\,\, 0 \,\,\, \\
0 & \mu_{2}^{2} &  \,\,\, 0  \,\,\, &  \,\,\, 0  \,\,\, \\
0 & 0 &  \,\,\, 0  \,\,\, &  \,\,\, 0  \,\,\,\\
0 & 0 & \,\,\, 0 \,\,\, &  \,\,\, 0 \,\,\,  \end{array}\right)\, ,\end{eqnarray}
with $\langle
F_{\Phi_{33}}\rangle  =  h\mu_{3}^{2}$ and 
$\langle F_{\Phi_{44}}\rangle  =  h\mu_{4}^{2}$. 
The vacuum energy of this minimum is 
$V_{+}=|h|^{2}(|\mu_{3}|^{4}+|\mu_{4}|^{4})$, where we have ordered 
$|\mu_{1}|>|\mu_{2}|>|\mu_{3}|>|\mu_{4}|>0$. 

We should ensure that there are no tachyonic directions around this vacuum. 
Along directions orthogonal 
to $f$ and $\tilde{f}$, the analysis of \cite{ISS} ensures that all 
mass-squareds are positive. Along the $f$ and $\tilde{f}$
directions themselves, diagonalizing their mass matrix yields\footnote{Assuming
$\kappa h$ to be real, without loss of generality.}
mass eigenstates $\xi_\pm=\frac{1}{\sqrt{2}} ({f}\pm e^{i\theta} \tilde{f}^*)$ 
where $\theta=Arg[\mu_3^2+\mu_4^2 ]$. Their mass-squareds are 
$m^2_{\xi_\pm}=m_f^2 \mp \kappa h |\mu_3^2+\mu_4^2 | $. 
The dominant contribution to the effective operator \eqref{eq:loop2},
\begin{equation}
W_{R}\, =\, \kappa\,\frac{g^2_{A}}{16\pi^2} \,\frac{Tr(\Phi)}{m_{R}}\,W_{A}^{\alpha}W_{\alpha}^{A},
\label{eq:loop3}
\end{equation}
comes from the loop with the lightest 
eigenstate $\xi_{-}$ circulating in the loop, and this determines $m_R$ in the denominator of
\eqref{eq:loop3}
to be 
\be
\label{eq:mr}
m_R\,=\,m_{\xi_{-}}\,=\,\sqrt{m_f^2 -\, h\kappa\,|\mu_3^2+\mu_4^2|} \, .
\ee
To maximize the contribution of the operator \eqref{eq:loop3}
to the gluino masses, as in subsection {\bf 3.1}, we
take $m_R\sim \mu \sim 100$ GeV and $h \kappa \gg 1$. This implies that 
$m_f \sim \sqrt{h \kappa}\, \mu $. 

The second candidate vacuum is characterized by 
\begin{eqnarray}
\label{second}
\langle f.\tilde{f}\rangle & = & \frac{h}{2\kappa}{(\mu_{3}^{2}+\mu_4^2)}\nonumber \\
\langle \Phi_{33}\rangle+\langle \Phi_{44}\rangle & = & -\frac{m_{f}}{\kappa}\nonumber \\
\langle \varphi_{i}\tilde{\varphi}_{j}\rangle & = & \left(\begin{array}{cccc}
\mu_{1}^{2}-\frac{1}{2}(\mu_{3}^{2}+\mu_4^2) & 0 & \,\,\, 0 \,\,\,  & \,\,\, 0 \,\,\, \\
0 & \mu_{2}^{2}-\frac{1}{2}(\mu_{3}^{2}+\mu_4^2) &  \,\,\, 0  \,\,\, &  \,\,\, 0  \,\,\, \\
0 & 0 &  \,\,\, 0  \,\,\, &  \,\,\, 0  \,\,\,\\
0 & 0 & \,\,\, 0 \,\,\, &  \,\,\, 0 \,\,\,  
\end{array}\right)
\end{eqnarray}
There are two vanishing $F_{\Phi_{ii}}$-terms, 
$\langle F_{\Phi_{11}}\rangle =\langle F_{\Phi_{22}}\rangle =0$, and 
two nonvanishing $F$-terms, \begin{eqnarray}
\langle F_{\Phi_{33}} \rangle= - \langle F_{\Phi_{44}} \rangle & = & h\frac{(\mu_{4}^{2}-\mu_{3}^{2})}{2}\, ,\end{eqnarray}
where the $\mu'$s are now ordered so that these are the least possible
$F$-terms. (Note that here we do not necessarily require 
$|\mu_{1}|>|\mu_{2}|>|\mu_{3}|>|\mu_{4}|$).
The vacuum energy is $V_{+\, {\rm{new}}}=\frac{|h|^{2}}{2}
|\mu_{3}^{2}-\mu_{4}^{2}|^{2}.$
Generally, 
this configuration breaks whatever symmetry $f$ and $\tilde{f}$ couple
to, so is certainly not a desirable vacuum state 
if $f$ and $\tilde{f}$ are charged under $SU(3)_{c}$.

Thus for the purposes of this paper we want the system to be
in the first vacuum, but the second vacuum 
\eqref{second} always has lower energy
since $V_{+}=|h|^{2}(|\mu_{3}|^{4}+|\mu_{4}|^{4})\geq \, \rm{Inf}
(\frac{|h|^{2}}{2}|\mu_{i}^{2}-\mu_{j}^{2}|^{2}$). Therefore one should
ensure that the decay time is sufficiently long.
The tunnelling rate from the first to the 
second vacuum is indeed suppressed due to a large separation between the two vacua in the
$f$- and in the $\Phi$-directions.
A simple estimate of the bounce action gives
\be S_E\sim 2\pi^2 \frac{\langle{\rm fields}\rangle ^4}{\Delta V}\approx\, \frac{4\pi^2}{\kappa^2} 
\ee 
where $\langle{\rm fields}\rangle$ denotes the separation in the field space between the
two vacua. In our case it is $\sim \sqrt{h/\kappa} \, \mu$ for both, the $f$- and the $\Phi$-directions.

 $S_E$ is easily made $\gtrsim 400$ to suppress tunnelling by choosing $\kappa \lesssim 0.3$.
Note that this conclusion would have 
been similar if we had stayed with $N_f=3$, the only difference being that 
the second vacuum would have been supersymmetric. With $N_f=4$ 
all perturbative vacua are metastable.

Thus we have demonstrated that the system can be trapped 
in the metastable vacuum of the first type where the effective 
operator $W_R$ in \eqref{eq:loop3}, is generated. 
This in turn gives Majorana masses $M_{\lambda_A}$  for the gauginos $\lambda_A$
({\it cf.} section {\bf 3.1}):
\be
M_{\lambda_A}\,=\, h\,\kappa\,\frac{g^2_{A}}{16\pi^2}\, \frac{\mu_3^2}{m_R}. 
\label{gaugino-mass2}
\ee
If we take
$h \kappa \,\sim \,16\pi^2/g^2 \gg 1$ we obtain $M_{\lambda} \sim \mu \sim 100 {\rm GeV}.$

\end{document}